\documentclass{article}

\usepackage{amsmath}
\usepackage{arxiv}
\usepackage{dsfont}
\usepackage{booktabs}
\usepackage{longtable}
\usepackage{multirow}
\usepackage{graphicx}
\usepackage[utf8]{inputenc}
\usepackage[english]{babel}
\usepackage{enumitem}
\usepackage{amsthm}
\usepackage[ruled,vlined]{algorithm2e}
\usepackage{accents}
\usepackage{graphicx}
\usepackage{blindtext}
\usepackage{csvsimple}
\usepackage{tikz}
\usetikzlibrary{shapes}
\usepackage{pgfplots}
	\pgfplotsset{compat=newest}
\usepackage{pgfplotstable}
\usepgfplotslibrary{fillbetween}
\usepackage{amsfonts}

\usepackage{bm}
\usepackage{mathtools}
\usepackage[utf8]{inputenc} 
\usepackage[T1]{fontenc}    
\usepackage{hyperref}       
\usepackage{url}            
\usepackage{booktabs}       
\usepackage{amsfonts}       
\usepackage{nicefrac}       
\usepackage{microtype}      
\usepackage{lipsum}		
\usepackage{graphicx}
\usepackage{natbib}
\usepackage{doi}
\newcommand{\fig}[4]{\begin{figure}[ht]\centering\includegraphics[width=#1\linewidth]{Figures/#2}\caption{#3}\label{#4}\end{figure}\\}
\usepackage{makecell}

\theoremstyle{definition}
\newtheorem{definition}{Definition}[section]

\theoremstyle{remark}

\title{Multi-output Gaussian processes for Inverse Uncertainty Quantification in Neutron Noise Analysis} 

\date{} 					

\author{ {Paul Lartaud}\\
	Ecole polytechnique (CMAP)\\
	CEA DAM Ile-de-France\\
	\small{\texttt{paul.lartaud@polytechnique.edu}} \\
	\And
	{Philippe Humbert} \\
	CEA DAM Ile-de-France\\
	\small{\texttt{philippe.humbert@cea.fr}} \\
	\And
	{Josselin Garnier} \\
	Ecole polytechnique (CMAP)\\
	\small{\texttt{josselin.garnier@polytechnique.edu}} \\
}

\renewcommand{\headeright}{}
\renewcommand{\undertitle}{}
\renewcommand{\shorttitle}{Multi-output Gaussian processes for inverse uncertainty quantification in neutron noise analysis}

\hypersetup{
pdftitle={MOGP_for_inverse_UQ},
pdfsubject={stat.CO},
pdfauthor={Paul Lartaud, Philippe Humbert, Josselin Garnier},
}

\begin{document}
\maketitle

\begin{abstract}
In a fissile material, the inherent multiplicity of neutrons born through induced fissions leads to correlations in their detection statistics. The correlations between neutrons can be used to trace back some characteristics of the fissile material. This technique known as neutron noise analysis has applications in nuclear safeguards or waste identification. It provides a non-destructive examination method for an unknown fissile material. This is an example of an inverse problem where the cause is inferred from observations of the consequences.
\newline However, neutron correlation measurements are often noisy because of the stochastic nature of the underlying processes. This makes the resolution of the inverse problem more complex since the measurements are strongly dependent on the material characteristics. A minor change in the material properties can lead to very different outputs. Such an inverse problem is said to be ill-posed.
\newline For an ill-posed inverse problem the inverse uncertainty quantification is crucial. Indeed, seemingly low noise in the data can lead to strong uncertainties in the estimation of the material properties.
\newline Moreover, the analytical framework commonly used to describe neutron correlations relies on strong physical assumptions and is thus inherently biased.
\newline This paper addresses dual goals. Firstly, surrogate models are used to improve neutron correlations predictions and quantify the errors on those predictions. Then, the inverse uncertainty quantification is performed to include the impact of measurement error alongside the residual model bias. 
\end{abstract}

\section*{Introduction}
Neutron noise analysis describes a set of methods which aims at identifying a fissile material based on observations of neutron correlations in the multiplying medium (\cite{feynman1956dispersion}). These methods can have applications in nuclear safeguards, criticality accident detections or waste identification (\cite{pazsit2007neutron}). Neutron correlations can be described analytically within a simplified framework known as the point model approximation (\cite{cifarelli1986models}). This model can be used to solve an inverse problem and evaluate characteristics of the medium such as its prompt multiplication. However, the point model relies on strong physical assumptions which leads to a systematic error in the predictions. The bias introduced by the point model is often disregarded and the uncertainty quantification of neutron noise techniques are not often considered. 
\newline The objective of this paper is to provide a robust inverse uncertainty quantification method which accounts for the bias introduced by the point model. The inverse problem is solved with a Bayesian approach as is done in several works in stochastic neutronics (\cite{verbeke2016neutron, verbeke2016stochastic}). Surrogate models are then used to replace and generalize the point model. The surrogate models are based on Gaussian processes (\cite{williams2006gaussian}), which are a flexible non-parametric regression tool. The surrogate models provide better predictions of neutron correlations and are computationally cheap to call. On top of this, they also yield covariances of the predictions. The predicted covariances can be introduced into the Bayesian resolution of the inverse problem to include the model uncertainties into the estimation of the posterior distribution of the input parameters.
\newline From the results obtained, the method presented in this paper is able to significantly improve neutron noise techniques while simultaneously providing a robust inverse uncertainty quantification.
\newline This paper provides a brief description of the point model and presents a summary of Gaussian Process Regression and its extension to multi-output problems. Then the general methodology is presented and tested on two different examples extracted from neutron multiplicity experiments.

\section{Point model framework}
In this section, a brief introduction to the point model approximation in neutron noise theory is presented.

\subsection{Neutron correlations}
In a fissile material, one neutron can induce a fission leading to the birth of more neutrons and so on. This process create correlations between the neutrons. The successive fissions can be described as branching processes, where each fission event is a node in a fission chain leading to the creation of more branches. \newline
Let us consider a set-up with a fissile material and a neutron detector. At some point, two neutron detections can be recorded simultaneously in the detector, or more specifically within the same time gate of temporal size $T$. This simultaneous double detection can either be accidental if the neutrons are independent, or correlated if they are not. Two neutrons are correlated if and only if they belong to the same fission chain. With this criteria, it is possible to make a distinction between true correlated double detections, and accidental double detections. \newline
Neutron noise techniques study the occurrence of true correlated double and triple detections in the the detectors.

\subsection{Point model assumptions}
The following assumptions are made in the point model framework. 
\begin{itemize}
    \item The medium is infinite, homogeneous, isotropic and subcritical. 
    \item Only fission and capture reactions occur in the material. They are governed respectively by the macroscopic cross sections $\Sigma_{f}$ and $\Sigma_{c}$
    \item Neutrons are monoenergetic.
    \item The source is either a spontaneous fission source, an $(\alpha ,n)$ source or a mix of the two. 
    \item Neutrons are detected by a neutron capture such that the total macroscopic capture cross section can be written $\Sigma_{c} = \Sigma_{d} + \Sigma_{p}$ where $\Sigma_{d}$ is the detection cross section, and $\Sigma_{p}$ the parasitic capture cross section.
\end{itemize}
These assumptions largely simplify the description of neutron correlations. An analytical description of neutron correlations is possible in this framework. The derivation of the point model equations is not provided here. Depending on the inputs and outputs considered, the point model equations can be found under different forms. Most of the time, Böhnel equations are used but in this paper the Feynman/Furuhashi framework is studied instead. 

\subsection{Böhnel equations}
The Böhnel equations establish a link between the material characteristics and neutron correlations. The outputs investigated are the single count rate $C_{1}$, the correlated double detections count rate $C_{2}$ and the correlated triple detections count rate $C_{3}$. \newline
The point model equations depend on the distribution of neutron multiplicity in induced and spontaneous fissions. The average number of neutrons produced per induced fissions is denoted $\overline{\nu}$. The second and third order factorial moments of the multiplicity distribution $\overline{\nu}_{2} = \overline{\nu (\nu-1)}$ and $\overline{\nu}_{3} = \overline{\nu (\nu-1)(\nu-2)}$ are introduced. \newline
The average number of neutrons for spontaneous fissions $\overline{\nu}_{s}$ and the factorial moments for spontaneous fissions  $\overline{\nu}_{2s}$ and $\overline{\nu}_{3s}$ are defined similarly.\newline
The input parameters in Böhnel framework are the leakage multiplication $M_{L}$, the leakage efficiency $\varepsilon_{L}$ defined as the number of counts per non-fissioning neutrons, and the fission source intensity $Q_{F}$ which is the number of spontaneous fission source events per second. Finally, the alpha ratio $\alpha_{R} = \frac{\overline{\nu}_{s} Q_{F}}{Q_{\alpha}}$ is the ratio of source neutrons produced by spontaneous fissions over the number of neutrons produced by $(\alpha, n)$.\newline
Based on the assumptions described in the previous sections, the point model equations can be derived.
\begin{equation}
    C_{1} = \varepsilon_{L} M_{L} \overline{\nu}_{s} Q_{F} (1 + \alpha_{R} )
\end{equation}
\begin{equation}
    \small
    C_{2} = \frac{\varepsilon_{L}^{2} M_{L}^{2} Q_{F}}{2} \left( \overline{\nu}_{2s} + \overline{\nu}_{2} \overline{\nu}_{s}(1 + \alpha_{R})\left(\frac{M_{L} -1}{\overline{\nu}-1} \right) \right)
\end{equation}
\begin{equation}
    C_{3} = \frac{\varepsilon_{L}^{3} M_{L}^{3} Q_{F}}{6}  \left(\overline{\nu}_{3s} + 3(1 + \alpha_{R})\overline{\nu}_{s}\overline{\nu}_{2}^{2}\left(\frac{M_{L} -1}{\overline{\nu}-1} \right)^{2} + ((1 + \alpha_{R})\overline{\nu}_{3}\overline{\nu}_{s} + 3 \overline{\nu}_{2s} \overline{\nu}_{2})\left(\frac{M_{L} -1}{\overline{\nu}-1} \right)  \right)
\end{equation}
\normalsize
A detailed derivation of the equations can be found in \cite{bohnel1985effect}.

\subsection{Feynman equations}
In Feynman/Furuhashi framework (\cite{furuhashi1968third}), the inputs considered are the prompt multiplication factor $k_{p}$, the detector efficiency (or Feynman efficiency) $\varepsilon_{F}$ defined as the number of counts per induced fissions, the source intensity $S$ defined as the number of source events (spontaneous fission or $(\alpha, n)$ reaction) per second, and $x_{s} = (1 + \alpha_{R})^{-1}$ which is the ratio of source neutrons produced per spontaneous fissions, over the total number of source neutrons. \newline 
The Diven factors of second and third order $D_{2}$ and $D_{3}$ for induced fissions are also introduced. They are defined as the second and third order reduced factorial moments of the multiplicity distribution. 
\begin{equation}
    D_{2} = \frac{\overline{\nu(\nu - 1)}}{\overline{\nu}^{2}} = \frac{\overline{\nu}_{2}}{\overline{\nu}^{2}} \ \ \ \ \ \ \ D_{3} = \frac{\overline{\nu(\nu - 1)(\nu -2)}}{\overline{\nu}^{3}} = \frac{\overline{\nu}_{3}}{\overline{\nu}^{3}}
\end{equation}
The Diven factors for the spontaneous fission multiplicity $D_{2s}$ and $D_{3s}$ are defined similarly. \newline
The outputs of the Feynman model are the average count rate $R$ and the second and third asymptotic Feynman moment $Y_{\infty}$ and $X_{\infty}$. \newline
The second and third order Feynman moments $Y(T)$ and $X(T)$ are defined for a given size $T$ of a detection window.
\begin{equation}
    Y(T) = \frac{2 N_{2C}(T)}{N_{1}(T)} \ \ \ \ \ \ X(T) = \frac{6 N_{3C}(T)}{N_{1}(T)} 
\end{equation}
where $N_{2C}(T)$ and $N_{3C}(T)$ are respectively the average number of double and triple correlated detections in a time window of size $T$ and $N_{1}(T)$ is the average number of count per time window. In this formalism the average count rate is the average number of counts divided by the time width. It does not depend on $T$.
\begin{equation}
    R = \frac{N_{1}(T)}{T} 
\end{equation}
The Feynman moments depend on the size of the time window considered. However, for $T$ large, $Y(T)$ (resp. $X(T)$) converges towards an asymptotic value $Y_{\infty}$ (resp. $X_{\infty}$). Indeed, in a  subcritical medium the fission chains have a limited lifetime. For $T$ much larger than the average fission chain lifetime, all the true correlations are encapsulated in the detection window. Thus, the Feynman moments reach an asymptotic value. \newline
Based on Böhnel equations, the equations for the Feynman/Furuhashi framework can be derived. The prompt reactivity $\rho = \frac{k_{p}-1}{k_{p}} < 0$ is introduced.  
\begin{equation}
    R = \frac{\varepsilon_{F} S \ \overline{\nu}_{s}}{- \rho \overline{\nu} \left( \overline{\nu}_{s} + x_{s} - \overline{\nu}_{s} x_{s} \right)} 
\end{equation}
\begin{equation}
    Y_{\infty} = \frac{\varepsilon_{F} D_{2}}{\rho^{2}} \left(1 - x_{s} \rho \frac{\overline{\nu}_{s} D_{2s}}{\overline{\nu} D_{2}} \right)
\end{equation}
\begin{equation}
    X_{\infty} = 3 \left(\frac{\varepsilon_{F} D_{2}}{\rho^{2}}\right)^{2} \left(1 - x_{s} \rho \frac{\overline{\nu}_{s} D_{2s}}{\overline{\nu} D_{2}} \right) - \frac{\varepsilon_{F}^{2} D_{3}}{\rho^{3}}^{2} \left(1 - x_{s} \rho \frac{\overline{\nu}_{s}^{2} D_{3s}}{\overline{\nu}^{2} D_{3}} \right)
\end{equation}

\subsection{Limitations}
The point model framework is based on strong physical assumptions. It is expected to yield biased predictions especially in cases where strong heterogeneities are present in the medium. \newline
A simple geometry is created to highlight this bias. Let us consider a metallic Pu spherical shell containing $94.0$ at.\% of $^{239}$Pu and $6.0$ at.\% of $^{240}$Pu, with at.\% referring to the atomic percentage of the isotope in the material. The shell has a density of $19.6040$ g.cm$^{-3}$ with internal and external diameter of respectively $7.014$ and $12.208$ cm. The internal region is void. The fissile region is surrounded by a borated polyethylene layer with density $1.00$ g.cm$^{-3}$ and $5.8$ at.\% of boron carbide B$_{4}$C, and thickness of $3.024$ cm. \newline
The average count rate and the Feynman moments are evaluated based on the method described later in section \ref{data_creation}. Similarly, the input parameters $\theta = (k_{p}, \varepsilon_{F}, S, x_{s})$ are recorded. Then, the point model predictions for the inputs $\theta$ are compared to the measured values of $(R, Y_{\infty}, X_{\infty})$. The results are summarized in Table \ref{pm_limitations}.
\begin{table}[ht]
    \renewcommand{\arraystretch}{1.}
    \centering
    \begin{tabular}{|c||c|c|c|} 
        \hline
         & $R$ & $Y_{\infty}$ & $X_{\infty}$ \\
        \hline
        MCNP6 simulated data       & $15120$ & $1.39$ & $5.52$ \\
        \hline
        Point model prediction & $16260$ & $1.29$ & $4.88$ \\
        \hline
        Relative deviation (\%)    & $7.5$   & $7.2$  & $11.6$ \\
        \hline
    \end{tabular}
    \caption{Point model bias for a test case with a strong heterogeneous configuration}
    \label{pm_limitations}
\end{table}\\
The point model does provide significant bias for all three outputs. The most prominent error is the error on the count rate $R$ because this is the least noisy output. Consequently, in the inverse problem resolution, a bias on $R$ is more penalizing than a bias on $X_{\infty}$ which is significantly more noisy and thus has a larger associated variance. \newline
The large bias observed in the point model impact the inference of the input parameters in the inverse problem. For such ill-posed problems, this bias can lead to significant errors in the estimation of $\theta$. For that reason, one of the objectives of this paper is to build a surrogate model able to improve predictions of $(R, Y_{\infty}, X_{\infty})$. The surrogate model should also include an estimation of the uncertainty of the predictions in order to include the residual model bias in the resolution of the inverse problem.

\section{Building a surrogate model}
Our objective is to replace the point model by a surrogate model. The surrogate model should be able to provide better predictions than the point model. For that purpose,  more input parameters are considered when building the surrogate model in order to account for neutron spectrum, parasitic absorption or leakage, which are not considered in the point model. \newline
A metric of the quality of the predictions is also expected from the surrogate model. Namely, the goal is to be able to provide a mean prediction for the output vector $(R, Y_{\infty}, X_{\infty})$ with its associated covariance. \newline
The covariance predictions are meant to be used directly in the Monte-Carlo Markov Chain methods to sample the posterior distribution of the input vector $\theta$ (that may be $\theta = (k_{p}, \varepsilon_{F} , S, x_{s})$ or that may include more input parameters). The aim is to include the model uncertainty in the model parameters. In this work, Gaussian processes are used as surrogate models.

\subsection{Scalar Gaussian processes}
\subsubsection{Introductory concepts}
Gaussian Process Regression is a non-parametric Bayesian regression method. It is a flexible tool for regression that is able to quantify uncertainties in the predictions. \newline
In this section, a brief description of Gaussian Process Regression (GPR) is given. This is not meant to be a thorough guideline but rather an introduction for beginners.
\begin{definition}
A Gaussian process is a collection of random variables, such that any finite subset follows a multivariate normal distribution. The distribution of a Gaussian process is completely defined by its mean function $m(\ \cdot \ )$ and covariance function $k(\ \cdot \ , \ \cdot \ )$. The Gaussian process formalizes the concept of distributions over functions. \newline
Let us first consider scalar Gaussian processes, with one real output, and multi-dimensional inputs. The input dimension is $I$. 
If $f$ is a Gaussian process with mean $m$ and covariance function $k$, it is denoted :
\begin{equation}
    f \sim \mathcal{GP}\left( m(x), k(x, x') \right)
\end{equation}

Assuming a Gaussian process model $f \sim \mathcal{GP}\left( m(x), k(x, x') \right)$ with scalar outputs, one can draw samples of the distribution of functions evaluated over $N$ input points represented by the matrix $\mathbf{X} = (X_{i})_{i \leq N}\in \mathds{R}^{N \times I}$. This sample vector $\mathbf{f} \in \mathds{R}^{N}$ follows a multivariate normal distribution.
\begin{equation}
    \mathbf{f} \sim \mathcal{N} \left( m(\mathbf{X}), K(\mathbf{X}, \mathbf{X}) \right)
\end{equation}
where $m(\mathbf{X}) \in \mathds{R}^{N}$ is the mean vector and $K(\mathbf{X}, \mathbf{X}) \in \mathds{R}^{N \times N}$ is the covariance matrix defined by $K(\mathbf{X}, \mathbf{X})_{i, j} = k(X_{i}, X_{j})$ for $i, j \leq N$. 
\end{definition}

\subsubsection{Predictions with Gaussian processes}
Let us consider a Gaussian process $f \sim \mathcal{GP}\left( m(x), k(x, x') \right)$. Let $\mathbf{X} \in \mathds{R}^{N \times I}$ and $\mathbf{y} \in \mathds{R}^{N}$ be respectively $N$ training inputs and outputs. Similarly, let $\mathbf{X'} \in \mathds{R}^{N' \times I}$ and $\mathbf{y'} \in \mathds{R}^{N'}$ be respectively $N'$ test inputs and outputs. \newline
Given the properties of Gaussian processes, the joint distribution of training and test outputs is :
\begin{align}
        \begin{pmatrix}
           \mathbf{f} \\
           \mathbf{f'} \\
         \end{pmatrix}
         \sim \mathcal{N}\left( 
        \begin{pmatrix}
           m(\mathbf{X}) \\
           m(\mathbf{X'}) \\
         \end{pmatrix},
        \begin{pmatrix}
           K(\mathbf{X}, \mathbf{X}) & K(\mathbf{X}, \mathbf{X'}) \\
           K(\mathbf{X'}, \mathbf{X}) & K(\mathbf{X'}, \mathbf{X'}) \\
         \end{pmatrix}
         \right)
\end{align}
The conditional distribution of the test outputs $\mathbf{f'}$ given $\mathbf{X}$, $\mathbf{X'}$ and $\mathbf{y}$ can then be obtained.
\begin{gather}\label{predictions}
    \mathbf{f'}|\mathbf{f}, \mathbf{X}, \mathbf{X'} \sim \mathcal{N}\left( \mu_{C}, K_{C} \right) \\
    \mu_{C} = m(\mathbf{X'}) + K(\mathbf{X}, \mathbf{X'})^{T} K(\mathbf{X}, \mathbf{X})^{-1} \left(\mathbf{y} - m(\mathbf{X}) \right) \\
    K_{C} = K(\mathbf{X'}, \mathbf{X'}) - K(\mathbf{X}, \mathbf{X'})^{T}K(\mathbf{X}, \mathbf{X})^{-1}K(\mathbf{X}, \mathbf{X'})
\end{gather}
where $\left(K(\mathbf{X}, \mathbf{X'})\right)_{i, j} = \left(K(\mathbf{X'}, \mathbf{X})^{T}\right)_{i, j}  = K(X_{i}, X'_{j})$ for $1 \leq i \leq N$ and $1 \leq j \leq N'$. \newline
This conditional distribution provides a way to predict the mean output from given input points $\mathbf{X'}$ as well as the covariance. The main interest of Gaussian process regression for our application is its ability to quantify the uncertainty on the predictions. \newline
For most situations, the available outputs often display noisy values such that observations are given by $\mathbf{y} = f(\mathbf{X}) + \mathbf{\varepsilon}$ with $\mathbf{\varepsilon} \sim \mathcal{N}(\mathbf{0}, \sigma_{obs}^{2}\mathcal{I}_{N})$. $\mathcal{I}_{N}$ refers to the identity matrix. The noise is assumed Gaussian and independent identically distributed. The covariance function is modified by adding a white noise kernel. 
\begin{equation}
    k_{noise}(X_{i}, X_{j}) = k(X_{i}, X_{j}) + \sigma_{noise}^{2} \delta_{i,j} \ \ \ \text{for } i, j \leq N
\end{equation}
where $\delta$ is the Kronecker symbol. Equation \ref{predictions} holds if $K(\mathbf{X}, \mathbf{X})$ is replaced by $K(\mathbf{X}, \mathbf{X}) + \sigma_{noise}^{2} \mathcal{I}_{N}$. Similarly, $\sigma_{noise}^{2} \mathcal{I}_{N'}$ is added to $K(\mathbf{X'}, \mathbf{X'})$ to predict the measurement noise. \newline
Predictions with Gaussian processes require the inverse of $K_{\sigma} = K(\mathbf{X}, \mathbf{X}) + \sigma_{noise}^{2} \mathcal{I}_{N}$. This inverse is obtained by a Cholesky decomposition which has good numerical stability properties. However, the matrix inversion has a complexity $O(N^{3})$. This is why GP regression does not scale well with very large dataset. In this case, the dataset size is around $1000$ which is sufficiently low to have acceptable training times. Exact GP regression can be carried out.  

\subsubsection{Choice of covariance functions}\label{cov_defintion}
The covariance function defines the regularity of the functions sampled from the Gaussian process. Different families of covariance functions exist and are used depending on the expected shape of the function to be learned. \newline
In this work, Radial Basis Function (RBF) (or squared-exponential) kernels and Matérn covariance kernels are used. 
\begin{definition}
Let $x$ and $x' \in \mathds{R}^{I}$. The anisotropic squared exponential covariance function $k_{SE}$ is a covariance function defined by :
\begin{equation}
    k_{SE}(x, x') = \sigma^{2} \prod\limits_{p=1}^{I} exp\left(-\frac{(x_{p}-x_{p}')^{2}}{2l_{p}^{2}} \right)
\end{equation}
The parameters $l_{p}$ are the correlations lengths for each input dimension $p$. The kernel is said to be isotropic if $l_{p}$ is independent of $p$.
\end{definition}
Squared exponential covariance functions are infinitely differentiable. As a result the corresponding Gaussian processes produce very smooth functions. The realizations of a Gaussian process with squared exponential covariance and mean zero are (almost surely) infinitely differentiable.  \newline
\begin{definition}
Let $x$ and $x' \in \mathds{R}^{I}$. Let $\nu \in \mathds{R}^{+}$. The Matérn class of anisotropic covariance functions $k_{Mat, \nu}$ is defined by :
\begin{equation}
    k_{Mat, \nu}(x, x') = \sigma^{2} \frac{2^{1-\nu}}{\Gamma(\nu)} \prod\limits_{p=1}^{I} \left( \frac{\sqrt{2\nu}|x_{p}-x_{p}'|}{l_{p}} \right)^{\nu} K_{\nu} \left( \frac{\sqrt{2\nu}|x_{p}-x_{p}'|}{l_{p}} \right)
\end{equation}
where $K_{\nu}$ is the modified Bessel function and $\Gamma(\nu)$ is the gamma function. \newline
\end{definition}
The Matérn functions are a broad class of covariance functions parametrized by $\nu$ which defines the regularity of the covariance kernel. When $\nu \longrightarrow + \infty$, the covariance function approaches the squared exponential. 
\begin{equation}
   \text{lim } k_{Mat, \nu}  \underset{\nu \rightarrow +\infty} = k_{SE} 
\end{equation}
The realizations of a Gaussian process with covariance function $k_{Mat, \nu}$ and mean zero are (almost surely) $n$-differentiable for $n < \nu$. The larger $\nu$, the smoother the Gaussian process. \newline
The Matérn covariances can be expressed as a product of an exponential and a polynomial of order $n$ for $\nu = n + 1/2$. \newline
The Matérn and squared-exponential covariance functions are widely used in Gaussian process regression. Since the functions to be learned are quite smooth here, Matérn 5/2 functions are chosen in the GP regressions. This class of covariance functions has been shown to provide the best performance for GP regression for this work. Matérn $3/2$ and squared-exponential kernels were also tested but provided lower performance. 

\subsubsection{Selection of hyperparameters}\label{hyperparameters}
In order to provide reasonable predictions for regression or classification problems, a Gaussian process has to be trained. The goal of the training phase is to choose the best parameters in the covariance kernels based on the training data. \newline
The common practice for selecting the hyperparameters is to find the values that maximize the marginal likelihood $p(\mathbf{y}|\mathbf{X})$. The marginal likelihood refers to the probability of the observations $\mathbf{y}$, integrated over all the possible function values $\mathbf{f}$ drawn from the Gaussian process. It is defined by :
\begin{equation}
    p(\mathbf{y}|\mathbf{X}) = \int p(\mathbf{y}|\mathbf{f}, \mathbf{X}) p(\mathbf{f}|\mathbf{X}) d\mathbf{f}
\end{equation}
From previous equations, $\mathbf{f}|\mathbf{X} \sim \mathcal{N}(\mathbf{0}, K(\mathbf{X}, \mathbf{X)})$ and $\mathbf{y}|\mathbf{f} \sim \mathcal{N}(\mathbf{f},  \sigma^{2}\mathcal{I}_{N})$. The log-marginal likelihood is thus given by :
\begin{equation}
    \text{log} p(\mathbf{y}|\mathbf{X}) = - \frac{1}{2} \mathbf{y}^{T}K_{\sigma}^{-1}\mathbf{y} - \frac{1}{2} \text{log}|K_{\sigma}| - \frac{N}{2}\text{log}(2\pi)
\end{equation}
with $K_{\sigma} = K(\mathbf{X},\mathbf{X}) + \sigma_{noise}^{2}\mathcal{I}_{N}$. The notation $|A|$ where $A$ is a square matrix refers to the determinant of the matrix $A$. The log-marginal likelihood is optimized using common optimization algorithms. In our case, the limited memory Broyden–Fletcher–Goldfarb–Shannon algorithm for bound constraint (also known as L-BFGS-B) is used (\cite{byrd1995limited}). The optimization algorithm is restarted $10$ times with different initial values for the hyperparameters. The optimal set of hyperparameters chosen is the one that provides the highest log-marginal likelihood of the $10$ iterations. With this approach, the risk of being stuck in a local optimum is reduced. \newline
Once the optimal set of hyperparameters is found, predictions can be made using equations \ref{predictions}.

\subsection{Multi-output Gaussian processes}\label{theory_mogp}
The framework described in the previous paragraphs assumes that the outputs are one-dimensional. However in this study, the outputs to be predicted are vectors $(R, Y_{\infty}, X_{\infty})$ with dimension $D=3$. For multi-output predictions, specific methods are required. Throughout the next paragraphs, the notations are extended for the multi-output case. Namely the outputs for the training set and the test set are now in the matrix form $\mathbf{y} \in \mathds{R}^{N \times D}$ and $\mathbf{y'} \in \mathds{R}^{N' \times D}$.

\subsubsection{Independent scalar Gaussian processes}
A first trivial approach is to train one Gaussian process for each output dimension. The Gaussian processes are trained independently from one another. This method is used in a first approach in this work though it has some strong flaws. \newline Independent training of the Gaussian processes means that the correlations between the outputs are not taken into account during the training phase, some information is lost. \newline
But more importantly, since the Gaussian processes are trained independently, the outputs must be assumed independent which means that one can only predict the variances of each output, but not the full covariance matrix. The objective of our method is to provide improved mean predictions but also the full covariance of the predicted outputs in order to include it into the inverse problem resolution. \newline
Thus, the goal is to build a multi-output Gaussian process model able to provide non-diagonal covariance predictions. In this work specifically, the outputs are strongly correlated with one another which makes this objective all the more important. \newline
Building a covariance kernel for a multidimensional GP is non-trivial since the covariance function has to remain positive definite. Several methods are investigated. 

\subsubsection{Linear Model of Coregionalization}
In order to build a multi-output covariance kernel, one possible method is to start off with independent scalar Gaussian processes and mix them with a transition matrix. This method is introduced in \cite{bonilla2007multi}. The matrix must be chosen to guarantee a positive definite covariance kernel. With this approach, it is possible to correlate the output channels while maintaining a positive definite covariance kernel. A brief description of this method is presented in the next paragraphs. \newline 
Let us consider $Q$ independent scalar Gaussian processes. For simplicity, only zero-mean Gaussian processes are considered.
\begin{equation}
    u_{q} \sim \mathcal{GP}\left(0, k_{q}(x, x') \right) \text{ for } 1 \leq q \leq Q
\end{equation}
These are called latent Gaussian processes. Now let us consider a real mixing matrix $W \in \mathds{R}^{D \times Q}$. Let $f_{d}$ be the output for channel $d$. It is obtained by the matrix product of $W$ and the vector of latent GPs. 
\begin{equation}
    f_{d}(x) = \sum\limits_{q=1}^{Q} w_{d,q}u_{q}(x)
\end{equation}
The covariance between two sets of inputs $\mathbf{X} \in \mathds{R}^{N \times I}$ and $\mathbf{X'} \in \mathds{R}^{N' \times I}$ for two channels $d$ and $d'$ can be calculated.
\begin{equation}
    Cov\left[f_{d}(\mathbf{X}), f_{d'}(\mathbf{X'}) \right] = \sum\limits_{q=1}^{Q} w_{d,q}w_{d', q} k_{q}(\mathbf{X}, \mathbf{X'})
\end{equation}
It is possible to flatten the multi-output vectors into a one-dimensional column.
\begin{equation}
    \mathbf{f(X)} = \left( f_{1}(X_{1}), ..., f_{1}(X_{N}), ..., f_{D}(X_{1}), ..., f_{D}(X_{N}) \right)^{T} \in  \mathds{R}^{DN}
\end{equation}
Now the covariance matrix can be written as a $DN \times DN'$ matrix using the Kronecker product $\otimes$. \newline
\begin{equation}
    \mathbf{K_{LMC}}(\mathbf{X}, \mathbf{X'}) = Cov\left[\mathbf{f(X)}, \mathbf{f(X')} \right] = \sum\limits_{q=1}^{Q} W_{\cdot, q} W_{\cdot, q}^{T} \otimes \mathbf{K_{q}}(\mathbf{X}, \mathbf{X'})
\end{equation}
where $W_{\cdot, q}$ is the $q$-th column of the matrix $W$ and $\mathbf{K_{q}}(\mathbf{X}, \mathbf{X'})$ is the covariance matrix obtained for kernel $k_{q}$ applied to the input sets $\mathbf{X}$ and $\mathbf{X'}$ . This defines a covariance kernel for the multi-output Gaussian process, where $\left(\mathbf{K_{LMC}}(\mathbf{X}, \mathbf{X'}) \right)_{((d-1)N + i, (d'-1)N' + j} = Cov\left[f_{d}(X_{i}), f_{d'}(X'_{j}) \right]$ for $d, d' \leq D$, $i \leq N$ and $j \leq N'$.\newline
The matrix inversion has now a complexity of $O\left(D^{3}N^{3} \right)$. Thus multi-output GPs are much more costly due to the size of the covariance matrix. For large output dimension, approximation methods such as Sparse Variational GP could be used (\cite{titsias2009variational}). In this work, the complexity remains reasonable and exact GP regression can be carried out. \newline
With this method, it is possible to build a multi-output Gaussian process able to provide the full covariance matrix of the predictions. 

\subsubsection{Convolutional Gaussian processes}
Similarly, one can build Convolutional Gaussian Processes by mixing the latent GPs with a convolution product instead of a matrix multiplication (\cite{alvarez2011computationally}). The output for the channel $d$ is built by the following relation :
\begin{equation}
    f_{d}(x) = \sum\limits_{q=1}^{Q} \int G_{d,q}(x-z)u_{q}(z) dz
\end{equation}
The functions $G_{d,q}$ represents filters in the convolution kernel. Let $G_{d, q}$ be a Gaussian filter with covariance $P_{d}^{-1}$, a positive definite matrix. Then the density $G_{d,q}(x)$ is the density of a Gaussian random variable $\mathcal{N}(0, P_{d}^{-1})$ with a multiplicative constant $S_{d,q}$. \newline
Let us consider the inputs $\mathbf{X} = (X_{i})_{i \leq N}\in \mathds{R}^{N \times I}$ and $\mathbf{X'} = (X'_{j})_{j \leq N'}\in \mathds{R}^{N' \times I}$. Let $k_{q}$ be the squared exponential covariance function corresponding to the $q$-th latent GP, with $q \leq Q$. It is written in a matrix form with $\Lambda_{q} = \text{diag}(l_{p, q}^{-2})_{p \leq I} \in \mathds{R}^{I \times I}$ a diagonal matrix containing the squared inverse correlation lengths $l_{p, q}^{-2}$ of the $q$-th latent GP for each input dimension $p \leq I$. The kernel $k_{q}$ also has a variance parameter $V_{q}$. Let us look at the covariance between $X_{i}$ for $i \leq N$ and $X'_{j}$ for $j \leq N'$. 
\begin{equation}
    k_{q}(X_{i}, X'_{j}) = V_{q}exp\left(-\frac{1}{2}(X_{i}-X'_{j})^{T}\Lambda_{q}(X_{i}-X'_{j}) \right)
\end{equation}
The convolution product in the expression of the covariance becomes tractable. 
\begin{equation}
    Cov(f_{d}(X_{i}), f_{d'}(X'_{j})) = \sum\limits_{q=1}^{Q} \frac{S_{d,q}S_{d', q}V_{q}}{(2\pi)^{I/2} |C_{d, d', q}|^{1/2}}
    exp\left(-\frac{1}{2}(X_{i}-X'_{j})^{T}C_{d, d', q}^{-1}(X_{i}-X'_{j}) \right)
\end{equation}
with $C_{d, d', q} = P_{d}^{-1} + P_{d'}^{-1} + \Lambda_{q}^{-1}$. It is possible to define the full covariance matrix.
\begin{equation}
    \mathbf{K_{CONV}}(\mathbf{X}, \mathbf{X'}) = \sum\limits_{q=1}^{Q} V_{q}
        \begin{pmatrix} 
            S_{1, q}S_{1, q} \mathbf{K_{1,1,q}}(\mathbf{X}, \mathbf{X'}) & \dots  & S_{D, q}S_{1, q} \mathbf{K_{1,D,q}}(\mathbf{X}, \mathbf{X'})\\
            \vdots & \ddots & \vdots\\
            S_{D, q}S_{1, q} \mathbf{K_{D,1,q}}(\mathbf{X}, \mathbf{X'}) & \dots  & S_{D, q}S_{D, q}\mathbf{K_{D,D,q}}(\mathbf{X}, \mathbf{X'})
        \end{pmatrix} 
\end{equation}
where $\mathbf{K_{d, d', q}}(\mathbf{X}, \mathbf{X'}) = \left( \mathbf{K_{d, d', q}}(X_{i}, X'_{j}) \right)_{i \leq N, \ j \leq N'}$ is given by : 
\begin{equation}
     \mathbf{K_{d, d', q}}(X_{i}, X'_{j}) = (2\pi)^{-I/2} |C_{d, d', q}|^{-1/2}exp\left(-\frac{1}{2}(X_{i}-X'_{j})^{T}C_{d, d', q}^{-1}(X_{i}-X'_{j}) \right)
\end{equation}
On top of this, a white noise kernel is added for each of the output channel. \newline
For convolutional Gaussian processes, the number of hyperparameters to optimize is much larger than for independent processes since the coefficients of the matrices $P_{d}$, $\Lambda_{q}$ and the scalar $S_{d,q}$ have to be learned. To simplify the training process, the Gaussian filters are chosen with diagonal covariance matrices $P_{d}$. This reduces drastically the number of hyperparameters. The main flaw of these convolutional GPs is that they might provide functions that are too smooth, due to the convolutional product with Gaussian filters and the use of RBF kernels. 

\subsection{Bias learning}\label{bias_learning}
Instead of learning directly the Feynman moments and the average count rate, it is possible to learn the disparities between the point model and the simulated data. This idea is adapted from \cite{kennedy2000predicting} where the low-fidelity code is the point model and the high-fidelity code is MCNP6 (or a real experiment).
\begin{equation}
    y_{true} = f_{PM}(\theta) + f_{GP}(\theta) + \varepsilon
\end{equation}
where $f_{PM}(\theta)$ are the point model predictions for input $\theta$ and $f_{GP}$ is the GP to be trained. \newline
Since the training data are not necessarily positive in this case, the Box-Cox transform cannot be used. Instead, the data is preprocessed using the Yeo-Johnson transform (\cite{yeo2000new}). \newline
However the point model equations require neutron multiplicity values $(\overline{\nu}, D_{2}, D_{3}, \overline{\nu}_{s}, D_{2s}, D_{3s})$ for induced and spontaneous fissions. For predictions, one does not have information on the neutron multiplicity parameters. Two solutions could be considered. Either the nuclear multiplicity parameters are included in the inputs of the forward model which increases the dimension of the input space to $13$. Or the nuclear multiplicity parameters are taken as the average over the training cases. However, this drastically limits the generalization of the surrogate model especially if other fissile isotopes are considered. For simplicity purposes, this second approach is used here. \newline
To account for systematic biases caused by the point model approximations, one can introduce an hyperparameter $\rho \in \mathds{R}$ and define a Gaussian Process $f_{GP}$ by weighting the contribution of the point model predictions with $\rho$.
\begin{equation}
    y_{true} = \rho f_{PM}(\theta) + f_{GP}(\theta) + \varepsilon
\end{equation}
The training of the GP works similarly except $\rho$ is identified as an hyperparameter to be tuned by maximizing the marginal log-likelihood. \newline
For the multi-output case, one can either take $\rho \in \mathds{R}$ as a scalar or $\mathbf{\rho} \in \mathds{R}^{D}$ as a vector and each of its component is weighting one output channel. The second approach is more flexible but introduces more hyperparameters. It is preferred in this work since the additional computational cost is not prohibitive. \newline
Overall, many different approaches have been presented for multi-output GP regression. These methods are all implemented and their performance are evaluated in section \ref{gp_performance}.

\section{General method}
\subsection{Extension of the point model}
The first objective is to extend the point model using surrogate models based on Gaussian processes. For that purpose, the input space needs to be extended to include more input parameters and to describe more accurately the underlying physical processes. \newline
First of all, the energy spectrum of the neutrons must be accounted for. Based on the two-group description commonly used in reactor physics, the spectral ratio $\varPhi$ is introduced. It is defined as the ratio of thermal flux over fast flux, taken in the surrounding medium of the object studied. 
\begin{equation}
    \varPhi = \frac{\phi_{th}}{\phi_{fast}}
\end{equation}
Next, parasitic absorptions should be included in the model description. In the point model, only the fissile material is considered while in practice neutron capture occurs in the moderating material as well (if any) or in reflectors. In order to account for this, a new input parameter $\varepsilon_{A}$ is defined, as the ratio of absorptions outside the fissile material, over induced fissions. 
\begin{equation}
    \varepsilon_{A} = \frac{\text{parasitic absorptions}}{\text{induced fissions}}
\end{equation}
Finally, the leakage is monitored by the ratio of outward over inward neutron current at the outermost layer of the MCNP6 model. In the training cases, this current ratio is taken at the concrete walls surrounding the object.
\begin{equation}
    J_{ratio} = \frac{J_{out}}{J_{in}}
\end{equation}
Overall, the input space is now $7$-dimensional with input parameters $\theta = \left(k_{p}, \varepsilon_{F}, S, x_{s}, \varepsilon_{A}, \varPhi, J_{ratio} \right)$. The MCMC sampling is more difficult since the dimension is higher. However, the additional parameters will be constrained using a simplified MCNP6 geometry that can be run quickly.

\subsection{Creating the dataset}\label{data_creation}
Sufficient training data must be fed to the surrogate models to construct efficient forward models. The training data are created using analog 3D Monte-Carlo simulations with the code MCNP6 (\cite{mcnp}). \newline
The unknown object to identify is a spherical fissile medium containing $^{239}$Pu and $^{240}$Pu and an inner void region of unknown diameter. It is surrounded by borated polyethylene, which serves as a moderator and a parasitic absorber. The object is placed in air at atmospheric pressure. The neutron detector consists of a cylindrical tube of $^{3}$He with CO$_{2}$ acting as a quench gas, and a cylindrical external shell of polyethylene used to slow down the incoming neutrons. The geometry is surrounded by a concrete layer modeling the walls of the room. The $^{3}$He detectors have a high efficiency but are only suited for the detection of thermal neutrons. Hence, they require a polyethylene outer layer to slow down the incoming neutrons. The main drawback is that the polyethylene washes out the temporal information of the slowing down process. This is not critical for this work, since only the asymptotic values of the Feynman moments are considered. However if their temporal dependence were to be included in the model, it would be more appropriate to consider detectors which do not bias the slowing down process, such as scintillators. \newline
The size, density and composition of the different regions are changed for each simulation. The fraction of boron in the borated polyethylene is changed for each case, and simulations are also performed without boron at all. The volumic source term is a mix of a spontaneous fission source and an $(\alpha, n)$ source located uniformly in the fissile region. The ratio of spontaneous fissions over $(\alpha, n)$ reactions is randomly changed for each training case in order to explore all the possible $x_{s}$. \newline
The prompt multiplication factor $k_{p}$ is obtained with a criticality calculation with $20$ inactive cycles and $100$ active cycles. The other parameters in the vector $\theta$ are obtained by tally measurements. The complete training set has a total of 1125 cases. \newline
The intrinsic physics of fission processes is not described. The neutron multiplicity distribution is modeled by the Terrel distribution (\cite{terrell1957distributions}).\newline
For each simulation the Feynman moments and the average count rate must be evaluated. All the neutron captures occurring in the detector are recorded in a time list file using the PTRAC command in MCNP6. This file is then post-processed to extract the Feynman moments. Since only the duration between the beginning of the neutron history and the time of detection is recorded in MCNP6, the birth times of the source neutrons must be sampled in the post-processing step. The source, whether it is a spontaneous fission source, an $(\alpha, n)$ source or a mix of both, is assumed to follow a Poisson statistics with intensity $S$. Then the birth instants of the source neutrons are defined by drawing samples from this Poisson statistics.  \newline
Based on the time list file obtained after the post-processing step, two main methods are used to evaluate the Feynman moments. They are described in the next paragraphs. \newline

\subsubsection{Sequential binning}\label{seq_bin_section}
In sequential binning, the numerical experiment is split into $W$ time windows of size $T$. In each window $i$, the number of counts $n_{i}(T)$ is recorded. From this, the simple moments of the detection statistics $M_{p}$ can be estimated. 
\begin{equation}
    \widehat{M_{p}} = \frac{1}{W} \sum\limits_{i=1}^{W} n_{i}(T)^{p}
\end{equation}
In the derivation of the point model equations, it can be seen that the Feynman moments are directly linked to the simple moments by the following relations. 
\begin{equation}
    R = \frac{M_{1}}{T}
\end{equation}
\begin{equation}
    Y = \frac{M_{2}}{M_{1}} - M_{1} - 1
\end{equation}
\begin{equation}
    X = \frac{M_{3}}{M_{1}} + 2(M_{1}^{2} + 1) - 3 \left( \frac{M_{2}}{M_{1}} + M_{2} - M_{1} \right)
\end{equation}
And thus, estimators of the Feynman moments can be obtained by replacing the simple moments by their estimators $\widehat{M_{p}}$ in the previous equations.  \newline
While the average count rate $R$ is independent of the size of the time window $T$, the Feynman moments are not. Once they are obtained for a given $T$, the same protocol can be repeated while merging $n \geq 1$ windows together to obtain estimators of $Y(nT)$ and $X(nT)$. \newline
Since the outputs of interest are the asymptotic Feynman moments $Y_{\infty}$ and $X_{\infty}$, $T$ should be large enough to reach the asymptotic state. However, the choice of a larger $T$ leads to less time windows $W$, and thus more noisy data due to accidental correlations. A compromise has to be made for the choice of $T$ between asymptoticity and noise. A time gate width of $T_{\infty} = 3 \ ms$ is used in this work. \newline
The sequential binning method can be used for numerical or practical experiment as long as a time list file is available. In this work, it is used to mimic practical evaluation of Feynman moments. All simulated data $\mathbf{y}$ from which the posterior distribution $p(\theta | \mathbf{y})$ is inferred are thus obtained using sequential binning. However, when sequential binning is used to obtain training data for the surrogate models, the inherent noise in the simulated data caused by accidental correlations reduces the performance of the surrogate models. Thus another method is used to evaluate the Feynman moments for that specific task.

\subsubsection{Filtered triggered binning}
The filtered triggered binning method allows to filter out accidental correlations using the knowledge provided by the numerical simulation. \newline
In this method, time windows are opened whenever a neutron is detected, and the history number of this neutron is kept in memory. For a given detection, the history number of the neutron is written in the time list file alongside the detection time. \newline
Then, the detections are recorded in the window if and only if the history number of the detected neutron is the same as the history number of the neutron that triggered the opening of the window. This means that only the correlated detections are recorded. An illustration of the method is shown in Figure \ref{trig_bin}\newline
\fig{1.0}{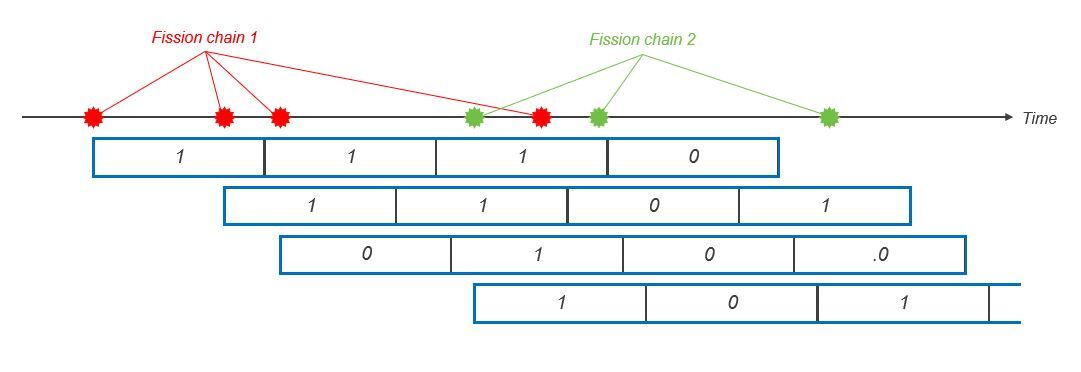}{Illustration of the filtered triggered binning method}{trig_bin}
\newline For a given window $i$, the recorded number of counts is $n_{i, tr}(T)$. The average number of double and triple counts which are coincidentally the second and third binomial moments of the detection distribution can then be obtained by the following estimators. 
\begin{equation}
    \widehat{m_{p}} = \frac{1}{N_{dec}} \sum\limits_{i=1}^{N_{dec}} \prod\limits_{k=0}^{p-2} \left( n_{i, tr}(T) - k \right)
\end{equation}
where $N_{dec}$ is the total number of detections and thus the total number of triggered windows. The Feynman moments can then be evaluated. 
\begin{align}
    \widehat{R} = \frac{N_{dec}}{T_{tot}} \\
    \widehat{Y} = 2\widehat{m_{2}} \\
    \widehat{X} = 6\widehat{m_{3}}
\end{align}
Filtered triggered binning is able to filter out the noise in the data but is only applicable because numerical simulations are used and the history number of the neutron is known. For a practical measurement, only sequential binning can be used. For that reason, filtered triggered binning is only used to create the training set for the surrogate models. Some inherent noise remains due to the stochastic nature of the MCNP6 simulations. These stochastic uncertainties are included in the GP covariance predictions. \newline
Once the surrogate models are trained, the inverse problem is solved with data obtained by sequential binning in order to mimic experimental data. Using filtered triggered binning data as training is a way to reduce the intrinsic noise in the dataset and improve the training of the surrogate models.

\subsection{Training the surrogate models}
\subsubsection{Dataset preprocessing}
Before training the surrogate models, the dataset needs to be pre-processed to improve the training performance. With the noise introduced by accidental correlations, the Feynman moments obtained for low $k_{p} < 0.6$ can become negative. These irregular data are removed from the dataset. \newline
Then, since the data distribution is far from Gaussian, a Box-Cox transform is applied (\cite{sakia1992box}). The Box-Cox transform is a non-linear transformation whose goal is to reshape the distribution into a standard normal distribution $\mathcal{N}\left(0, \mathcal{I} \right)$. It requires positive data. 
\begin{equation}
B(y, \lambda) =
\left\{
	\begin{array}{ll}
		\frac{\left(y^{\lambda} - 1 \right)}{\lambda}  & \mbox{if } \lambda \neq 0 \\
		\log(y) & \mbox{if } \lambda = 0
	\end{array}
\right.
\end{equation}
The parameter $\lambda$ is fitted to match a standard normal distribution by maximising a log-likelihood between the data and the standard normal distribution. \newline

\subsubsection{GP performance comparison}\label{gp_performance}
Before training, the dataset is split into a training set and a test set with the same size. The training set contains around 560 cases. \\ 
Different metrics are used to evaluate the performance on the test set. The Mean Absolute Error (MAE), Mean Squared Error (MSE) and Mean Absolute Percentage Error (MAPE) are used here. Let $y_{d,j}$ be the $j$-th test output for the output channel $d$, corresponding to inputs $\theta_{j}$ for $j \leq N'$. Let $\overline{f_{d}(\theta_{j})}$ be the mean surrogate model prediction for inputs $\theta_{j}$ and for the output channel $d$. Then for the output channel $d$ the metrics are defined by : 
\begin{equation}
    MAE_{d} = \frac{1}{N'} \sum\limits_{j=1}^{N'} \left|y_{d,j} - \overline{f_{d}(\theta_{j})} \right|
\end{equation}
\begin{equation}
    MSE_{d} = \frac{1}{N'} \sum\limits_{j=1}^{N'} \left( y_{d,j} - \overline{f_{d}(\theta_{j})} \right)^{2}
\end{equation}
\begin{equation}
    MAPE_{d} = 100 \times \frac{1}{N'} \sum\limits_{j=1}^{N'}\left| \frac{ y_{d,j} - \overline{f_{d}(\theta_{j})}}{y_{d,j}} \right|
\end{equation}
The robustness of the uncertainty quantification is also investigated. The coverage probability in the estimated $2\sigma$ confidence interval is provided. It is defined as the ratio of test outputs lying inside the $2\sigma$ confidence interval predicted by the surrogate model. The theoretical value should be $95.45 \%$. \newline
In a first approach, the GPs are used to learn directly the observations $(R, Y_{\infty}, X_{\infty})$. The methods investigated are the multi-output independent GPs, the Linear Model of Coregionalization (LMC) and  the Convolutional GPs described in section \ref{theory_mogp}. Their performance are displayed in Table \ref{gp_perf}.
\begin{table}[ht]
    \renewcommand{\arraystretch}{1.}
    \centering
    \begin{tabular}{|c||c|c|c|c|} 
        \hline
        Independent GPs & MAE & MSE & MAPE & Coverage prob. ($2 \sigma$) \\ 
        \hline
        Count Rate     & $1.98 \times 10^{2}$  & $1.10 \times 10^{5}$  & $0.95 \%$ & $95.3 \%$ \\
        \hline
        Second Feynman & $2.93 \times 10^{-2}$ & $3.84 \times 10^{-3}$ & $2.56 \%$ & $94.2 \%$  \\
        \hline
        Third Feynman  & $1.25 \times 10^{0}$  & $4.71 \times 10^{1}$  & $9.67 \%$ & $93.9 \%$ \\
        \hline
        \hline
        LMC - 2 latent GPs & MAE & MSE & MAPE & Coverage prob. ($2 \sigma$) \\ 
        \hline
        Count Rate     & $2.24 \times 10^{2}$  & $1.18 \times 10^{5}$  & $1.11 \%$ & $92.2 \%$ \\
        \hline
        Second Feynman & $3.77 \times 10^{-2}$ & $1.01 \times 10^{-2}$ & $2.85 \%$ & $94.0 \%$  \\
        \hline
        Third Feynman  & $1.19 \times 10^{0}$  & $4.25 \times 10^{1}$  & $9.73 \%$ & $94.6 \%$ \\
        \hline
        \hline
        LMC - 3 latent GPs & MAE & MSE & MAPE & Coverage prob. ($2 \sigma$) \\ 
        \hline
        Count Rate     & $2.20 \times 10^{2}$  & $1.09 \times 10^{5}$  & $1.10 \%$ & $93.3 \%$ \\
        \hline
        Second Feynman & $3.85 \times 10^{-2}$ & $9.22 \times 10^{-3}$ & $3.08 \%$ & $90.5 \%$  \\
        \hline
        Third Feynman  & $1.33 \times 10^{0}$  & $5.96 \times 10^{1}$  & $10.0 \%$ & $92.3 \%$ \\
        \hline
        \hline
        Convolutional GPs & MAE & MSE & MAPE & Coverage prob. ($2 \sigma$) \\ 
        \hline
        Count Rate     & $1.77 \times 10^{3}$  & $4.86 \times 10^{7}$  & $5.25 \%$ & $98.4 \%$ \\
        \hline
        Second Feynman & $4.62 \times 10^{-1}$ & $8.26 \times 10^{0}$ & $7.76 \%$ & $97.9 \%$  \\
        \hline
        Third Feynman  & $4.10 \times 10^{1}$ & $1.34 \times 10^{5}$  & $18.3 \%$ & $97.7 \%$ \\
        \hline
    \end{tabular}
    \caption{Performance of the Gaussian processes surrogate models on the test set}
    \label{gp_perf}
\end{table}
\newline
The convolutional GPs do not provide satisfying predictions. This is likely linked to the use of Gaussian filters and covariance kernels which hinder the GP to model accurately the data. Because of their poor performance, convolutional GPs are not studied in the next paragraphs. \newline
On the other hand, both Linear Coregionalization GPs and Independent GPs provide satisfying mean predictions and 1D coverage probability. However the independent GPs are expected to yield poor predictions for the 2D confidence ellipses.  Thus, the coverage probability for the 2D confidence ellipses are also investigated. Figure \ref{2D_coverage} displays the coverage probability for each couple of outputs and for the different surrogate models built. The coverage probability is shown for different levels of confidence.\newline
\fig{1.}{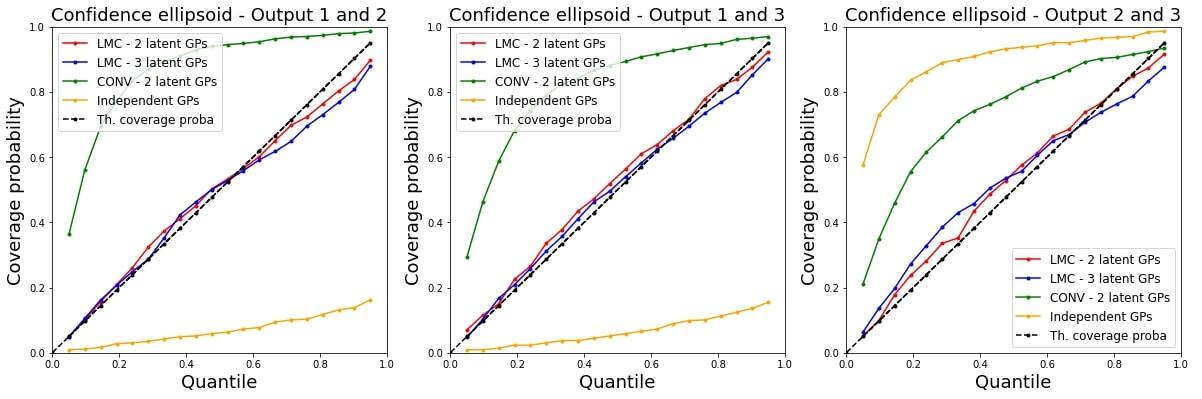}{2D Coverage probability for different levels of confidence obtained for the surrogate models: LMC with 2 (red) and 3 (blue) latent GPs, convolutional GP (green) and independent GPs (yellow)}{2D_coverage} \newline
From this figure, one can see that using independent GPs for the output channels does not yield robust covariance predictions. Besides, convolutional GPs tend to overestimate the variance regardless of the confidence levels required. On the other hand, LMC provides robust covariance predictions with 2 or 3 latent GPs. \newline
Overall, Linear Coregionalization GPs provide excellent predictions for the average count rate and the Feynman moments with robust covariance predictions. For that particular case, LMC with 2 latent GPs is the best-performing surrogate model, though its performance are comparable to LMC with 3 latent GPs. \newline
In a second approach, the GPs learn the disparities with the point model, first with fixed bias $\rho = 1$ (Table \ref{gp_perf_pm_disparities}) and then with non-fixed $\rho$ (Table \ref{gp_perf_pm_with_rho}). 
\begin{table}[ht]
    \renewcommand{\arraystretch}{1.}
    \centering
    \begin{tabular}{|c||c|c|c|c|c|} 
        \hline
        Independent GPs & MAE & MSE & MAPE & Coverage prob. ($2 \sigma$) \\ 
        \hline
        Count Rate     & $2.02 \times 10^{2}$  & $1.31 \times 10^{5}$  & $0.83 \%$ & $95.7 \%$ \\
        \hline
        Second Feynman & $5.88 \times 10^{-2}$ & $1.05 \times 10^{-1}$ & $2.78 \%$ & $93.1 \%$  \\
        \hline
        Third Feynman  & $1.06 \times 10^{0}$  & $3.93 \times 10^{1}$  & $8.95 \%$ & $94.2 \%$ \\
        \hline
        \hline
        LMC - 2 latent GPs & MAE & MSE & MAPE & Coverage prob. ($2 \sigma$) \\ 
        \hline
        Count Rate     & $2.05 \times 10^{2}$  & $1.14 \times 10^{5}$  & $0.94 \%$ & $93.9 \%$ \\
        \hline
        Second Feynman & $3.52 \times 10^{-2}$ & $7.93 \times 10^{-3}$ & $2.74 \%$ & $93.9 \%$  \\
        \hline
        Third Feynman  & $1.08 \times 10^{0}$  & $3.42 \times 10^{1}$  & $8.51 \%$ & $94.8 \%$ \\
        \hline
        \hline
        LMC - 3 latent GPs & MAE & MSE & MAPE & Coverage prob. ($2 \sigma$) \\ 
        \hline
        Count Rate     & $2.06 \times 10^{2}$  & $1.13 \times 10^{4}$  & $0.96 \%$ & $93.1 \%$ \\
        \hline
        Second Feynman & $3.30 \times 10^{-2}$ & $7.47 \times 10^{-3}$ & $2.54 \%$ & $93.1 \%$  \\
        \hline
        Third Feynman  & $1.07 \times 10^{0}$  & $3.61 \times 10^{1}$  & $8.63 \%$ & $96.0 \%$ \\
        \hline
    \end{tabular}
    \caption{Performance of the bias learning Gaussian processes with $\rho = 1$, on the test set}
    \label{gp_perf_pm_disparities}
\end{table}\\

\begin{table}[ht]
    \renewcommand{\arraystretch}{1.}
    \centering
    \begin{tabular}{|c||c|c|c|c|c|} 
        \hline
        Independent GPs & MAE & MSE & MAPE & Cov. prob. ($2 \sigma$) & $\rho$ \\ 
        \hline
        Count Rate     & $2.00 \times 10^{2}$  & $1.17 \times 10^{5}$  & $0.93 \%$ & $93.0 \%$ & $1.038$ \\
        \hline
        Second Feynman & $3.44 \times 10^{-2}$ & $8.23 \times 10^{-3}$ & $2.38 \%$ & $95.8 \%$ & $0.950$ \\
        \hline
        Third Feynman  & $9.23 \times 10^{-1}$  & $2.74 \times 10^{1}$  & $9.19 \%$ & $95.7 \%$ & $0.964$ \\
        \hline
        \hline
        LMC - 2 latent GPs & MAE & MSE & MAPE & Cov. prob. ($2 \sigma$) & $\rho$ \\ 
        \hline
        Count Rate     & $2.08 \times 10^{2}$  & $1.27 \times 10^{5}$  & $0.96 \%$ & $92.6 \%$ & $1.022$ \\
        \hline
        Second Feynman & $3.34 \times 10^{-2}$ & $6.83 \times 10^{-3}$ & $2.65 \%$ & $94.2 \%$ & $0.962$ \\
        \hline
        Third Feynman  & $1.12 \times 10^{0}$  & $4.09 \times 10^{1}$  & $8.44 \%$ & $95.4 \%$ & $0.923$ \\
        \hline
        \hline
        LMC - 3 latent GPs & MAE & MSE & MAPE & Cov. prob. ($2 \sigma$) & $\rho$ \\ 
        \hline
        Count Rate     & $2.08 \times 10^{2}$  & $1.22 \times 10^{5}$  & $0.96 \%$ & $92.2 \%$ & $1.032$ \\
        \hline
        Second Feynman & $3.83 \times 10^{-2}$ & $8.51 \times 10^{-3}$ & $3.11 \%$ & $91.8 \%$ & $0.941$ \\
        \hline
        Third Feynman  & $1.18 \times 10^{0}$  & $4.56 \times 10^{1}$  & $8.70 \%$ & $94.0 \%$ & $0.964$ \\
        \hline
    \end{tabular}
    \caption{Performance of the bias learning Gaussian processes with $\rho \neq 1$, on the test set}
    \label{gp_perf_pm_with_rho}
\end{table}
One can see that for all the surrogate models considered, the bias learning yields overall better performance. Fixing $\rho = 1$ appears to be the best performing method. When $\rho$ is not fixed, the optimization step is made more difficult due to the three additional parameters (for multi-output GPs), which can in turn impact the performance of the surrogate model. \newline
Overall, this study shows that LMC, with 2 or 3 latent GP, is able to provide robust surrogate models with good prediction performance. The surrogate models can be further improved  by learning the bias with the point model instead of directly learning the outputs $(R, Y_{\infty}, X_{\infty})$. For the rest of this work the surrogate model used is the LMC GP with 2 latent GPs.

\subsubsection{Sensitivity analysis}
Once the GPs are trained, a sensitivity analysis can be performed to evaluate the impact of the additional parameters $(\varepsilon_{A}, \varPhi, J_{ratio})$. The objective is to understand whether or not the additional parameters affect the predictions. \newline
The sensitivity analysis is done by evaluating the Sobol indices of total order $S_{T, p}$ for input $p$. Let $\mathbf{X}$ be a random variable of dimension $I$ representing the inputs, and $Y$ a scalar output random variable. The inputs coordinates are assumed independent. Let $\mathbf{X}_{-p} = (X_{1},  ..., X_{p-1}, X_{p+1}, ..., X_{I})$ be the vector of input variables without the $p$-th coordinate. Then the total order Sobol index for the dimension $p$ is given by the relative fraction of the variance explained by the interaction of the $p$-th coordinate with all the other inputs. 
\begin{equation}
    S_{T, p} =  \frac{\mathds{E}\left[Var(Y| \mathbf{X}_{-p}) \right]}{Var(Y)}
\end{equation}
The Sobol indices can be estimated with Monte-Carlo estimators, by the pick-freeze method (\cite{marrel2009calculations}). \newline
The total order Sobol indices are shown in Table \ref{table_sobol}. As expected, the main input of importance is $k_{p}$ for the Feynman moments and $S$ plays a significant role in the evaluation of the count rate $R$, while having very limited influence on the Feynman moments. One can observe that each of the additional inputs has an impact on at least one of the outputs that is similar to $x_{s}$. Thus, the additional inputs are indeed significant in the evaluation of the outputs, though their respective impacts are much more limited than for $(k_{p}, \varepsilon_{F}, S)$. 

\begin{table}[ht]
    \renewcommand{\arraystretch}{1.}
    \centering
    \begin{tabular}{|c||c|c|c|} 
        \hline
        Total order Sobol & $R$ & $Y_{\infty}$ & $X_{\infty}$ \\ 
        \hline
        $k_{p}$           & $3.72 \times 10^{-1}$ & $6.99 \times 10^{-1}$ & $8.16 \times 10^{-1}$ \\
        \hline
        $\varepsilon_{F}$ & $6.44 \times 10^{-1}$ & $4.54 \times 10^{-1}$ & $5.21 \times 10^{-1}$ \\
        \hline
        $S$               & $1.00 \times 10^{-1}$ & $1.48 \times 10^{-4}$ & $3.02 \times 10^{-4}$ \\
        \hline
        $x_{s}$           & $4.30 \times 10^{-3}$ & $2.82 \times 10^{-3}$ & $1.39 \times 10^{-3}$ \\
        \hline
        $\varepsilon_{A}$ & $5.59 \times 10^{-4}$ & $5.25 \times 10^{-4}$ & $1.30 \times 10^{-3}$ \\
        \hline
        $\varPhi$         & $2.29 \times 10^{-5}$ & $1.39 \times 10^{-3}$ & $3.22 \times 10^{-3}$ \\
        \hline
        $J_{ratio}$       & $4.55 \times 10^{-3}$ & $2.00 \times 10^{-4}$ & $4.03 \times 10^{-4}$ \\
        \hline
    \end{tabular}
    \caption{Total order Sobol indices estimated with pick-freeze method, with $10000$ MC samples}
    \label{table_sobol}
\end{table}

\subsection{Adaptive MCMC}
The sampling of the posterior distribution $p(\theta | \mathbf{y})$ of the parameter $\theta$ given simulated data $\mathbf{y}$ can be conducted directly with Bayes' theorem by choosing a fine mesh of the parameter space and by evaluating the likelihood in every point of the mesh. In the framework of the point model, the likelihood can be directly calculated, however this method is very cumbersome because it requires a fine discretization of the $4$-dimensional parameter space. It becomes even more difficult to implement as the number of dimensions increases. Working with the surrogate models instead of the point model would require the discretization of a $7$-dimensional space, which is very costly in terms of memory and running time. \newline
An alternative way of sampling the posterior distribution is to use \textit{Monte-Carlo Markov Chain} methods. The MCMC methods allow to create random samples whose empirical distributions converge towards the target distribution. Besides they only require knowledge of the target density within a multiplicative constant.  \newline
Markov chains are stochastic processes where one state only depends on the previous state. A Markov chain is said to be ergodic for a distribution $\pi$ if the expectation of a function with respect to this distribution (assuming it exists) can be approximated by the empirical average of this function on the states of the chain. \newline
More specifically, let $X$ be a random variable following a law $\pi$. Let $(X_{i})_{i \in \mathds{N}}$ be an ergodic Markov chain for the distribution $\pi$. Then for any function $f$ such that $\mathds{E}_{\pi} \left[|f \right|] = \int_{\mathcal{X}} |f(x)|\pi(dx) < + \infty $
\begin{equation}
    \mathds{E}_{\pi} \left[f \right] = \lim_{N \rightarrow + \infty} \frac{1}{N}\sum\limits_{i=1}^{N} f(X_{i})
\end{equation}
The goal of the MCMC algorithms is to build an ergodic Markov chain for a given target distribution. For this work, the target distribution to sample is the posterior distribution of the input parameters given some observations $p(\theta |  \mathbf{y})$. The ergodic property allows to estimate different quantities of interest for the distribution such as its mean, its variance, all the moments, the quantiles, the probabilities of being in a given interval and so on. 

\subsubsection{Adaptive Metropolis}
One of the simplest MCMC algorithm is the Metropolis-Hastings algorithm (\cite{metropolis1953equation, hastings1970}). Its goal is to create an ergodic Markov chain whose invariant distribution is the target distribution $\pi$. \newline
The Metropolis-Hastings algorithm is robust, but has some limitations in the case of very degenerate target distributions as it is the case in this work. A degenerate distribution is a distribution whose support lies mainly on a subspace (or a manifold) of the parameter space, whose dimension is strictly lower. It can be for example a curve or a plane in the $7D$ parameter space. This definition of degeneracy is not rigorous. More precisely, a degenerate distribution is rigorously defined as a distribution whose support has Lebesgue measure equal to zero. However the notion of degeneracy is considered in this work as a practical limitation to MCMC methods and not as a formal mathematical definition.\newline
In such a case, since the support of the distribution is thin, most of the candidate points in Metropolis-Hastings tend to miss the support and be rejected and consequently the acceptance rate is close to $0$. If the proposal covariance is adjusted to reach a higher acceptance rate, the distribution is not properly sampled and the chain stays around the same spot. \newline
One way to correct this is to adapt the covariance matrix of the proposal distribution, in order to draw candidate points closer to the support of the distribution. \newline
The \textit{Adaptive Metropolis} (AM) algorithm presented in this section is more thoroughly detailed in \cite{haario2001adaptive}.\newline
Let $\pi$ be the target distribution and $\theta_{0}$ the initial point of the chain. In Metropolis-Hastings, at each iteration $n \geq 1$, a candidate point is sampled with a proposal distribution $\widehat{\theta_{n}} \sim q(\cdot | \theta_{n-1}) \sim \mathcal{N}\left(\theta_{n-1}, \mathcal{C} \right)$ where the proposal covariance $\mathcal{C}$ is usually chosen diagonal. The idea of the AM algorithm is to adapt the covariance of the proposal by estimating the empirical covariance of the previously accepted points of the chain. In this method, the proposal distribution is of the form $\widehat{\theta_{n}} \sim \mathcal{N} \left( \theta_{n-1}, \mathcal{C}_{n-1} \right)$, but the covariance is modified at each step to match the empirical covariance of the points of the chain. 
\begin{equation}
    \mathcal{C}_{n} = s \times Cov\left(\theta_{0}, ... , \theta_{n} \right)\text{ with }Cov\left(\theta_{0}, ... , \theta_{n} \right) = \frac{1}{n} \sum\limits_{i=0}^{n} \left(\theta_{i}-\overline{\theta_{n}}\right) \left(\theta_{i}-\overline{\theta_{n}}\right)^{T}
\end{equation}
with $\overline{\theta_{n}} = \frac{1}{n+1} \sum\limits_{i=0}^{n} \theta_{i}$. The scalar $s$ is a scaling parameter that needs to be tuned to reach the desired acceptance rate. \newline
The direct calculation of the empirical covariance is cumbersome when the chain becomes long. A recursive formula is preferred to evaluate $\mathcal{C}_{n}$ and $\overline{\theta_{n}}$. 
\begin{equation}
    \overline{\theta_{n+1}} = \frac{1}{n+2} \left(\theta_{n+1} + (n+1)\overline{\theta_{n}} \right)
\end{equation}
\begin{equation}\label{recursive_cov}
    \mathcal{C}_{n+1} = \frac{n-1}{n} \ \mathcal{C}_{n} + \frac{s}{n} \left(n \ \overline{\theta_{n}} \ \overline{\theta_{n}}^{T} - (n+1) \ \overline{\theta_{n+1}} \ \overline{\theta_{n+1}}^{T}   + \theta_{n+1} \ \theta_{n+1}^{T} \right)
\end{equation}
These recursive formulas help speed up the calculation of the covariance. In practice, it is advised to add a small term of the form $\epsilon \mathcal{I}_{d}$ with $\epsilon > 0$ in order to guarantee the matrix stays positive definite. Indeed, numerical approximations can lead to a degenerate covariance matrix, which can be problematic for the sampling of the candidate points. \newline
The acceptance rate of the candidate points in MCMC methods is a key factor to monitor. For Metropolis-Hastings, it was shown that the optimal acceptance rate was roughly $0.234$ in high dimension (\cite{gelman1997weak}). The scaling factor of the proposal covariance must be tuned in order to reach an acceptance rate close to this value. A naive search for a good scaling factor can be performed but it is also possible to dynamically change the scaling factor to reach a target acceptance rate. One solution explored in \cite{haario2001adaptive}, is to multiply the covariance at each step by a factor $r_{n}$ defined as :
\begin{equation}
    r_{n} = \text{exp} \left( \alpha_{n} - \alpha_{target} \right)
\end{equation}
where $\alpha_{target}$ is the target acceptance rate and $\alpha_{n}$ is the current acceptance rate. This method is implemented in the AM algorithm in our work.  \newline
The adaptation of the covariance matrix is not started directly from the beginning, but rather after a certain number of accepted points $n_{0}$ is reached in order to make sure the empirical covariance is calculated on enough points. $n_{0}$ is set at $500$ in our case. \newline
The Adaptive Metropolis algorithm (AM) is described below.
It can be shown that the AM algorithm retains the ergodic property (\cite{andrieu2006ergodicity}). \newline
\begin{algorithm}[H]
\SetAlgoLined
  \KwResult{Sampling of the target distribution $\pi$}
  Choose the chain starting point $\theta_{0}$ and desired length $K$\;
  Set the start of adaptation $n_{0}$ \;
 \While{Chain length $<K$}{
  Generate candidate $\widehat{\theta}$ from proposal distribution $q(\widehat{\theta}|\theta_{n})$\;
  Evaluate the acceptance probability $\alpha(\widehat{\theta},\theta_{n}) = min \left\{1,\frac{\pi(\widehat{\theta}) q(\theta_{n}|\widehat{\theta})}{\pi(\theta_{n}) q(\widehat{\theta}|\theta_{n})} \right\}$\;
  Generate $u \sim \mathcal{U}[0,1]$ from a uniform distribution on $[0,1]$\;
  \eIf{$\alpha(\widehat{\theta},\theta_{n}) > u$}{
   Add $\widehat{\theta}$ to the chain $\theta_{n+1}=\widehat{\theta}$\;
   \eIf{Number of iterations $\geq n_{0}$}{
   Adapt the proposal covariance with equation \ref{recursive_cov} \;
   }{
   Keep the same proposal covariance
   }
   }{
   Add $\theta_{n}$ to the chain $\theta_{n+1}=\theta_{n}$\;
  }
 }
 \caption{Adaptive Metropolis}
\end{algorithm}
The AM algorithm is more suited to degenerate probability distributions as the proposal distribution aligns with the distribution support and the candidate points are closer to the target distribution. \newline
A variant of this algorithm is the Adaptive Proposal (\cite{haario1999adaptive}) where the covariance adaptation is performed locally, using the $H$ previous points instead of all the previous points. However, for this method the invariant measure is biased with respect to the target distribution. For this reason, the AM algorithm is used.

\subsubsection{Coupling MCMC and surrogate model predictions}\label{coupling}
The forward model used to predict the outputs $y$ for given inputs $\theta$ is found in the evaluation of the posterior density $p(\theta | \mathbf{y})$ in the AM algorithm, when evaluating the acceptance probability $\alpha(\widehat{\theta},\theta_{n})$. \newline
Let us consider $N$ independent observations, with each including the count rate and the second and third Feynman moments $\mathbf{y} = (y_{i})_{i \leq N} = ((R, Y_{\infty}, X_{\infty})_{i})_{i \leq N}$. Using Bayes' theorem, the posterior can be written as the product of a prior distribution $p(\theta)$ and a likelihood $L(\mathbf{y}| \theta)$ which is the probability distribution of the observations $\mathbf{y}$ given the inputs $\theta$. \newline
\begin{equation}
    p(\theta | \mathbf{y}) \propto p(\theta)L(\mathbf{y}| \theta)
\end{equation}
It is assumed the $N$ independent observations are given by $y = f(\theta) + \varepsilon$ where $f$ is a forward model used to predict the outputs and $\varepsilon \sim \mathcal{N}\left( \mathbf{0}, \mathbf{C}_{obs} \right)$. The forward model can be for example the analytical point model or a GP surrogate model. Then, the acceptance probability in MCMC can be easily evaluated since the likelihood is Gaussian and the observations $\mathbf{y} = (y_{i})_{1 \leq i \leq N}$ are independent.
\begin{equation}\label{likelihood}
    L(\textbf{y}|\theta) \propto \text{exp}\left(-\frac{1}{2}\sum_{i = 1}^{N} (y_{i}-f(\theta))^{T} \mathbf{C}_{obs}^{-1} \ (y_{i}-f(\theta)) \right)
\end{equation}
This requires to know the covariance of the noise $\mathbf{C}_{obs}$. This covariance is not simply a diagonal matrix because the three output channels are strongly correlated. The covariance could be estimated simply by taking the empirical covariance of the observations however this is not very efficient because the number of independent observations is typically around $10$. \newline
Instead, a bootstrap method is used to evaluate this covariance matrix (\cite{efron1994introduction}). The number of bootstrap samples is set to $10000$. The outputs are strongly correlated as expected with $Corr(Y_{\infty}, X_{\infty}) \simeq 0.9$ for example. \newline
Hence the target density is known within a multiplicative constant. The MCMC sampling can now be performed using either the point model, or a surrogate model for better predictions. \newline
However, even though surrogate models based on GP regression do perform better than the point model for predicting the outputs, they also come with uncertainties in their predictions. In order to have a robust uncertainty quantification method, these model uncertainties must be accounted for in the posterior distribution sampling. \newline
In order to include the model uncertainties, the likelihood is modified. It is assumed that the model errors and the noise errors of the data are independent. Thus the statistical model, for a GP surrogate model can be rewritten as $y = \overline{f(\theta}) + \varepsilon_{model}(\theta) + \varepsilon_{noise}$ where the noise is $\varepsilon_{noise} \sim \mathcal{N}\left( \mathbf{0}, \mathbf{C}_{obs} \right)$ and the model error is $\varepsilon_{model}(\theta) \sim \mathcal{N}\left( \mathbf{0}, Cov\left[f(\theta) \right] \right)$. $\overline{f(\theta})$ is the mean prediction of the GP and $Cov\left[f(\theta) \right]$ is the covariance prediction of the GP at input point $\theta$. \newline
Since the model error and noise are assumed independent the likelihood can be modified to include the model error (\cite{higdon2004combining}). However, the simulated data are not independent anymore since they are all linked by the same model error $\varepsilon_{model}(\theta)$. \newline
The residuals $\mathbf{R}(\theta) = \left(y_{i} - \overline{f(\theta}) \right)_{i \leq N} \in \mathds{R}^{DN}$ and the total covariance $\mathbf{C}_{tot}(\theta) \in \mathds{R}^{DN \times DN}$ are introduced.
\begin{equation}
    \mathbf{C}_{tot}(\theta) = 
    \begin{pmatrix} 
        \mathbf{C}_{obs} + Cov\left[f(\theta) \right] & \dots & Cov\left[f(\theta) \right]\\
        \vdots & \ddots & \vdots\\
        Cov\left[f(\theta) \right] & \dots & \mathbf{C}_{obs} + Cov\left[f(\theta) \right]
    \end{pmatrix} 
\end{equation}
The model error can be included in the MCMC sampling by the mean of a modified likelihood.
\begin{equation}\label{mod_likelihood}
    L(\textbf{y}|\theta) \propto \frac{1}{\sqrt{\left|\mathbf{C}_{tot}(\theta) \right|}}  \text{exp}\left(-\frac{1}{2} \mathbf{R}(\theta)^{T} \mathbf{C}_{tot}(\theta)^{-1} \ \mathbf{R}(\theta) \right)
\end{equation} 
The evaluation of the likelihood is more cumbersome now since a matrix inversion and a determinant calculation must be performed at each iteration in MCMC, yet this is still manageable for low output dimension. This new likelihood does include the model uncertainty in the sampling of the posterior distribution $p(\theta | \mathbf{y})$. It is used with the AM algorithm to sample the posterior distribution. In the next section, this method is used on a test case from the ICSBEP Handbook. 

\section{Application to the BERP sphere ICSBEP Benchmarks}
The method presented in this paper is tested on two examples. The first one is taken from the ICSBEP Handbook (\cite{briggs2003international}) and the second one is based on the same fissile object with a different moderating material. Simplified descriptions of the two cases are built in MCNP6 and used to simulate neutron correlation data. From this, the posterior distribution of the parameters $\theta$ is sampled and compared to the MCNP6 values.
\subsection{MCNP6 modeling}
\subsubsection{Copper-reflected plutonium sphere}
This first example is taken from the set of experiments FUND-NCERC-PU-HE3-MULT-003 of the ICSBEP Handbook, experiment n°1. The experiment is a measurement of neutron correlations on the BERP sphere, a metallic Pu sphere with a mean diameter of $7.5876$ cm and an average density of $19.6039$ g.cm$^{-3}$. The sphere is surrounded by a SS-304 cladding consisting of two hemispheres. For simplicity, the hemispheres are modeled as a single spherical shell of SS-304 with inner and outer diameter of $7.65556$ and $7.71652$ cm respectively.\newline
The sphere is surrounded by a single layer of a spherical shell of copper reflector with inner and outer diameter of $7.7978$ and $10.1600$ cm respectively.\newline
An aluminum structure was supporting the BERP ball. However, to simplify the MCNP6 model this structure is not included. The expected prompt multiplication factor evaluated should be slightly lower than for the practical experiment due to the absence of the reflections on the support structure. \newline
In the experiment, two NOMAD detectors were placed on each side of the BERP ball. In the simplified model, only one detector is modeled. The detectors have no influence on each other as shown in the benchmark. The NOMAD detectors consist in a series of 15 $^{3}$He tubes with active dimensions of $2.46 \times 38.1$ cm, set in a polyethylene moderating block. This design is simplified in the model used in this work. The detectors are replaced by a single cylindrical $^{3}$He tube. The diameter of the tube is chosen to have the same active volume as for the real experiment. The diameter of the tube is thus $9.52$ cm. Similarly the thickness of polyethylene is set to $5.00$ cm. It is chosen to have the same total polyethylene mass as in the experiment.  \newline
The detector region is filled with a mixture of $^{3}$He with $2$ at.\% of CO$_{2}$ acting as a quench gas, where at.\% refers to the atomic percentage of an element. The pressure is set to $10.13$ bars. \newline
The source intensity is chosen in the post-processing step for the sampling of the neutron birth instants. The source intensity is chosen to be $S = 132582 \text{ events}.s^{-1}$ and with $x_{s} = 0.969$. \newline
All nuclear data, including effective cross sections and neutron multiplicities, are considered known. In some works, the uncertainties in the nuclear data are propagated to the neutron transport simulations. This is technically feasible with the methodology described in this paper but that would further complicate the statistical models involved. For simplification purposes, the nuclear data are thus considered known. Besides, in this work, the main source of uncertainty comes from the limited number of neutron detections, especially for the third Feynman moment. The impact of the nuclear data uncertainties is expected to be negligible in comparison. \newline
The detector efficiency, the second Feynman moment $Y(T)$ and the prompt multiplication estimated by the neutron correlation observations are also provided (\cite{hutchinson2019validation}). The time gate width is set to $T = 2048 \  \mu s$ in the benchmark. These values are compared to the ones obtained with the simplified MCNP6 model built to create our own Feynman observations. The second Feynman, as well as $k_{p}$ and $\varepsilon_{F}$ are shown in Table \ref{comparison_ICSBEP_MCNP_copper} with their respective standard deviation. \\

\begin{table}[ht]
    \renewcommand{\arraystretch}{1.}
    \centering
    \begin{tabular}{|c||c|c|c|} 
        \hline
         & $Y(T)$ & $k_{p}$ & $\varepsilon_{F}$ \\
        \hline
        Experiment & $0.339 \pm 0.008$ & $0.8306 \pm 0.0009$ & $0.0216$ \\
        \hline
        Simplified MCNP6 & $0.39 \pm 0.01$  & $0.8279 \pm 0.002$ & $0.0182 \pm 0.0002$ \\
        \hline
    \end{tabular}
    \caption{Comparison between the ICSBEP experiment and the simplified MCNP6 model for the copper-reflected plutonium sphere}
    \label{comparison_ICSBEP_MCNP_copper}
\end{table}

Overall, the simplified MCNP6 model built is close to the benchmark model and the experiment. Some disparities arise due to the simplifications made. They can be linked to the removal of the aluminum support plate for example. Our goal here is not necessarily to exactly retrieve the benchmark results but rather to show the improvements brought by the method presented in this paper with an application on a well-documented experiment. \\

\subsubsection{Polyethylene-reflected BERP sphere}
In the training set, the fissile material is surrounded by borated polyethylene and thus the copper-reflected example displays some disparities with the actual training examples. \newline
These disparities might bring a bias in the posterior distribution of $\theta$. For this reason, a second example is tested where this time the BERP ball is surrounded by polyethylene. This example is taken from \cite{mattingly2009polyethylene}. The methodology is similar. \newline
The polyethylene shell has internal and external diameter of $7.798$ and $15.418$ cm and a density of $0.95$ g.cm$^{-3}$. \newline
The fissile object is unchanged and thus the source intensity is the same $S = 132582 \text{ events}.s^{-1}$. \newline
The benchmark does not provide the prompt multiplication factor or the detector efficiency. The values obtained with the MCNP6 model are $k_{p} = 0.8967 \pm 0.0004$ and $\varepsilon_{F} = 0.0118 \pm 0.0001$. The second Feynman moment for $T = 2048 \ \mu s$ is $Y_{exp}(T) = 0.75 \pm 0.05$ which is close to the value obtained with the simplified model used $Y_{MCNP}(T) = 0.814 \pm 0.014$. \newline
As in the previous case, small disparities are observed between the experiment and the simplified MCNP6 model used in this work. 

\subsection{Posterior distribution sampling}
The two cases described before are used to create neutron correlation observations from which the posterior distribution $p(\theta | \mathbf{y})$ is sampled with Adaptive Metropolis. 

\subsubsection{Sampling with the point model}\label{pm_sampling_section}
A first approach is to use only the point model as a forward model. The likelihood is then given by equation \ref{likelihood}. \newline
This method does not provide a way to quantify the bias of the point model. Only the observation covariance $\mathbf{C}_{obs}$ is added into the MCMC sampling. \newline
The data are simulated by running $10$ independent MCNP6 simulations, with $2 \times 10^{6}$ neutron histories for each. The Feynman moments are evaluated using the sequential binning method described in section \ref{seq_bin_section}. The time gate width is set to $T_{\infty} = 3 \ ms$. \newline
Adaptive Metropolis is used to sample the posterior distribution. The sampling is performed over $5 \times 10^{6}$ iterations, parallellized on $5$ CPUs. The total running time of MCMC is $5$ minutes. The first $2000$ iterations are not recorded in the chain, in order to make sure the chain has reached its stationary distribution. This is called the burn-in phase. The target acceptance rate is set to $0.12$. It is lower than the usual $0.234$ because the target distribution is very degenerate and far from a regular Gaussian target. The covariance adaption is started after $5000$ iterations to ensure enough points are recorded in the chain so that the empirical covariance is not too degenerate. A regularization term is added to the proposal covariance to avoid reaching non-positive definite matrix because of numerical instabilities. \newline
The prior $p(\theta)$ for $\theta = \left( k_{p}, \varepsilon_{F}, S, x_{s}\right)$ is set to a uniform distribution on a broad domain $\mathcal{D}$.
\begin{equation}
    \mathcal{D} = [0.70, 0.95] \times [0.001, 0.100] \times [1 \times 10^{5}, 2 \times 10^{5}] \times [0.4, 1.0]
\end{equation}
The choice of the prior should be non-informative. This prior does not assume good prior knowledge of the input parameters. An even broader domain could be used but this would require more training cases to completely cover the prior domain. Since exact GP regression is used in this work, the prior domain is kept small enough even though this means the prior is not completely non-informative. \newline
One could also use Jeffreys prior (\cite{jeffreys1946invariant}) which is a non-informative prior designed to be invariant under reparametrization of the forward model. In the point model framework, Jeffreys prior is analytically tractable. However, it was shown in a previous work than the effect of the prior for the MCMC sampling does not affect significantly the posterior distribution as long as enough observations are provided (\cite{lartaud2022}).\newline
The maximum a posteriori $\theta_{map}$ obtained is not really close to the real values. 
\begin{equation}
    \theta_{cu} = (0.856, 0.0128, 151500, 0.68)
\end{equation}
The 2D marginal densities for $(k_{p}, \varepsilon_{F}, S)$ are shown in Figure \ref{2D_marg_pm_sampling_copper}.
\newline One can see that the real values of the parameters lie far outside the actual posterior distribution. This is a consequence of the bias of the point model which is not accounted for in this method.
\fig{1.0}{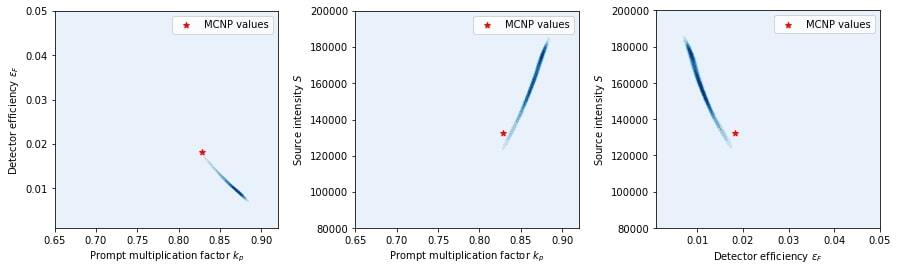}{2D marginal densities of the posterior distribution sampled by MCMC with only the point model, for the copper-reflected sphere}{2D_marg_pm_sampling_copper}
\newline The same calculation are performed for the polyethylene case. The 2D marginal densities are displayed in Figure \ref{2D_marg_pm_sampling_poly}.
\fig{1.0}{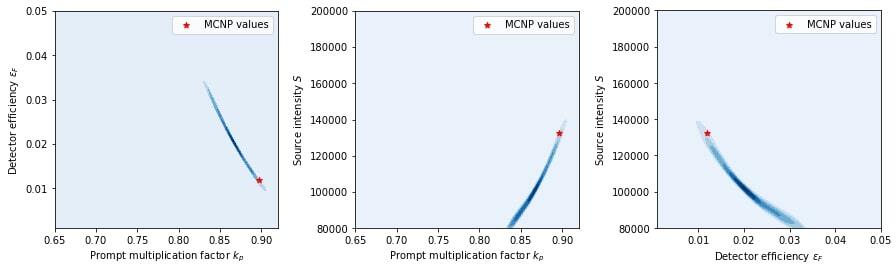}{2D marginal densities of the posterior distribution sampled by MCMC with only the point model, for the polyethylene-reflected sphere}{2D_marg_pm_sampling_poly}
The maximum a posteriori $\theta_{poly}$ itself does not provide good predictions once again. 
\begin{equation}
    \theta_{poly} = (0.860, 0.0197, 108300, 0.83)
\end{equation}
For the polyethylene case, the real values lie closer to the sampled posterior distribution. Indeed, with the polyethylene the neutrons are thermalized such that the assumption of monoenergetic neutrons is more reasonable than for the copper-reflected case. \newline
Overall, the posterior distribution sampling with only the point model as a forward model does not provide robust uncertainty quantification nor a reliable prediction on the input parameters $\theta$. This can be improved by using the surrogate models.

\subsubsection{Prior constraints on the domain}
In order to improve the sampling of the posterior distribution, the surrogate model is coupled to the MCMC sampling as described in section \ref{coupling}. \newline
The input parameter space is now $7$-dimensional which makes the MCMC sampling much more difficult. To counterbalance the higher dimension, more restrictive bounds are placed on the prior. \newline
The parameters of interest are mainly $(k_{p}, \varepsilon_{F}, S, x_{s})$. The three additional parameters have less impact on the predictions. They can be constrained either by direct practical measurement, or by numerical simulations with MCNP6. In this work, a few 1D MCNP6 calculations are run to estimate these parameters. For each simulation, the parameters are evaluated by tally measurements. It is assumed the external diameter of the fissile region is known and fixed. This information can be obtained by $\gamma$-spectroscopy for example. The internal diameter is changed for each simulation.\newline
The size of the 1D spherical regions are chosen to preserve the total mass compared to the 3D model. Since the detector is all around the fissile object, a correction based on the real solid angle $\Omega$ of the detector as seen by the fissile region is applied. Only $\Omega / 4\pi$ neutrons are actually recorded in the $^{3}$He detector. \newline
The bounds for the uniform prior for $(\varepsilon_{A}, \varPhi, J_{ratio})$ are then chosen as the minimal and maximal values obtained in the 1D simulations. The prior $p(\theta)$ is a uniform distribution on the extended domain $\mathcal{D}_{ext}$. 
\begin{equation}
    \mathcal{D}_{ext} = \mathcal{D} \times [3\times 10^{-3}, 5 \times 10^{-3}] \times [10^{-4}, 10^{-3}] \times [1.7, 1.9]
\end{equation}

\subsubsection{Surrogate model sampling}
The surrogate model used is the bias learning Linear Coregionalization GP with 2 latent processes and non-fixed $\rho \neq 1$. This surrogate model displayed the best performance, though comparable to the other LMC surrogate models. The posterior distribution sampling is performed similarly as in section \ref{pm_sampling_section}. The MCMC sampling is done over $5 \times 10^{6}$ iterations once again, with a total running time of $40$ minutes. The running time is slightly longer since the modified likelihood evaluation in equation \ref{mod_likelihood} is more computationally demanding. \newline
\fig{1.0}{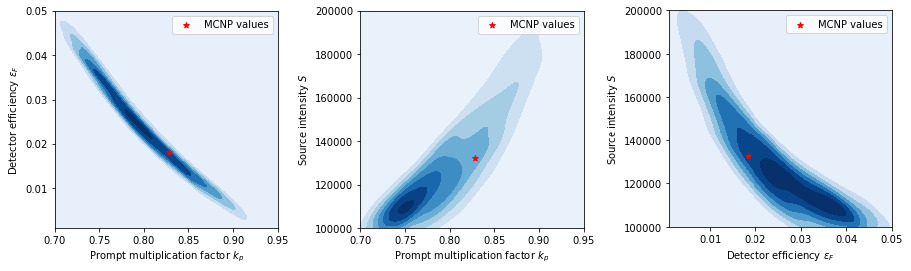}{2D marginal densities of the posterior distribution sampled with the bias learning LMC2 surrogate model, for the copper-reflected case}{2D_marg_surrogate_sampling_copper}\newline
The marginal densities for the copper and polyethylene reflected Pu sphere are displayed respectively in Figure \ref{2D_marg_surrogate_sampling_copper} and Figure \ref{2D_marg_surrogate_sampling_poly}. The maxima a posteriori for the copper $\theta_{cu}$ and the polyethylene $\theta_{poly}$ are also given. 
\begin{equation}
    \theta_{cu} = (0.806, 0.0222, 136000, 0.617)
\end{equation}
\begin{equation}
    \theta_{poly} = (0.865, 0.0228, 118300, 0.74)
\end{equation}
\fig{1.0}{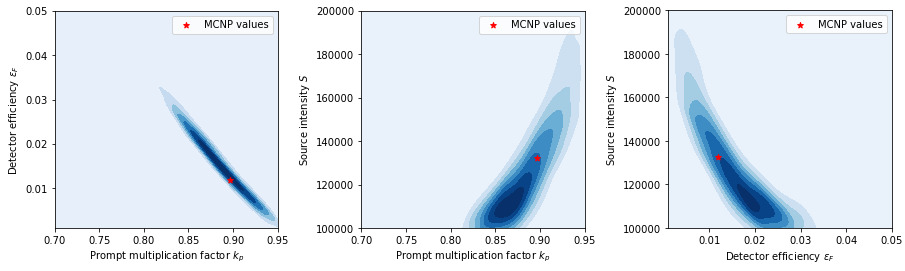}{2D marginal densities of the posterior distribution sampled with the bias learning LMC2 surrogate model, for the polyethylene-reflected case}{2D_marg_surrogate_sampling_poly}
\newline The posterior distributions sampled are much broader because the model error is now included. For both cases, the theoretical points are in the distribution support. The maxima a posteriori are slightly more precise than in section \ref{pm_sampling_section} but do not provide precise predictions on their own. The estimation of $x_{s}$ is especially difficult. \newline
The training cases used a polyethylene moderator around the sphere, which means that the copper-reflected case differs from the training set used. Yet the real values are within the support of the distribution. The surrogate models are flexible enough to provide reasonable predictions on cases different from the training set. \newline
The methodology presented in this paper allows for a robust inverse uncertainty quantification while being flexible in its use. As remarked above, the posterior distributions are quite broad due to the significant model errors in the surrogate models (of course, another reason is the rather limited amount of information given by the neutron correlation observations considered in the test cases). The global performance of the method for a given data set could be improved by reducing the model error. This could be achieved by using more involved surrogate modeling approaches as we discuss in the next subsection. That would allow to obtain narrower posterior distributions for real-world applications in nuclear safeguards or waste identification for example.

\subsection{Further improvements}
The surrogate models described in this paper are based on homoscedastic Gaussian Processes. The white noise kernel added to account for the noise in the training data is assumed constant over the whole range of data. \newline
However, from the observations made in this work, the noise increases with $k_{p}$.
\fig{0.7}{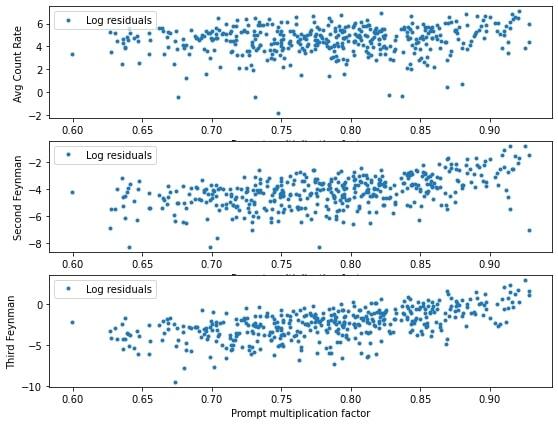}{Plot of the log residuals (in absolute value) for the test case as a function of $k_{p}$ for the LMC2 GP regression}{plot_residuals}
In Figure \ref{plot_residuals}, the log-residuals (in absolute value) of the GP regression are plotted as a function of $k_{p}$ for the test case. The residuals appear to be larger for large $k_{p}$. This plot illustrates the heteroscedastic nature of the data used in this work. \newline
In order to improve this methodology, heteroscedastic GPs could be created. For example, one can assume the variance in the training follows a parametric trend $\sigma_{noise}^{2}(x) = f_{\mathbf{\beta}}(x)$. Then the parameters $\mathbf{\beta}$ can be included in the hyperparameters selection described in section \ref{hyperparameters}. Other methods use a second GP as a surrogate model for the noise variance itself (\cite{kersting2007most}). \newline
The methodology is also compatible with other types of surrogate models as long as they are able to provide covariance predictions. Other supervised learning techniques could be applied to this problem. In this paper, the choice of the surrogate model was motivated by the simplicity and the flexibility of GP regression. But more complex techniques could improve the posterior distribution sampling. \newline
In this paper, the objective was not to model real-world measurements as accurately as possible, but rather to build a methodology able to solve the inverse problem and provide robust uncertainties on the predictions.
In order to apply this methodology to real-world measurement, one would require better performance for the surrogate models. This could be achieved by adding inputs for the description of the energy spectrum, or the slowing-down process. Besides, the surrogate models could benefit from a more detailed description of the fission dynamics, as is done with the fission models FREYA and FIFRELIN for example (\cite{verbeke2018correlated}).
Beyond neutron measurements, it is also possible to obtain information from other types of measurements (gamma spectroscopy and X-ray for instance). This information could be taken into account in the prior distribution of the proposed method.

\section{Conclusion}
\label{sec:conclusion}
Overall the methodology presented in this paper provides a framework for robust uncertainty quantification in neutron noise analysis. The extension of the point model with the help of surrogate models based on Gaussian processes allows for better predictions in the forward model and more consistent uncertainty quantification in the inverse problem resolution. \newline
The application to the two test cases has highlighted the robustness of the method for uncertainty quantification. Yet, the trained surrogate models still suffer from large variances which hinders precise estimation of fissile mass based on neutron correlation observations. Two lines of work are identified to improve the practical application of this methodology to real-world problems in nuclear safeguards or waste drum identification. First of all, the surrogate models can be improved with the help of heteroscedastic Gaussian Processes or other supervised learning techniques. On the other hand, one could also provide more information to the surrogate models to accurately model realistic neutron correlation measurements, such as better descriptions of slowing-down processes and fission dynamics. \newline
Despite the possible improvements aforementioned, the methodology presented in this paper allows for robust uncertainty quantification for inverse problems resolution in nuclear safeguards, coupled with affordable computational resources, while improving the standard approach based on the analytical point model description.

\pagebreak
\bibliographystyle{unsrtnat}                                              
\bibliography{bibliography}

\end{document}